# A Possible Model of Noise EnhancedVisual Perception in Human Vision

Ajanta Kundu
Applied Nuclear Physics Division
Saha Institute of Nuclear Physics
1/AF Bidhannagar, Kolkata, India
ajanta.kundu@saha.ac.in

Sandip Sarkar [#]
Applied Nuclear Physics Division
Saha Institute of Nuclear Physics
1/AF Bidhannagar, Kolkata, India
sandip.sarkar@saha.ac.in

*Abstract*—We demonstrate, through simulation, that a simple centre surround receptive field of vision is capable of exhibiting stochastic resonance. We also show that this could be used to model the nature of contrast sensitivity enhancement of human vision, through stochastic resonance, observed in psychophysical experiments.

I. INTRODUCTION

Stochastic resonance (SR) is a phenomenon whereby small amount of additive noise can significantly enhance the performance of a non-linear signal processing system. The concept of stochastic resonance was first proposed by Nicolas, [1] and Benzi [2], to explain the near periodicity of the ice-ages, which coincides with periodic variation of earth's orbital eccentricity though this periodic force was too weak to cause such an abrupt change in earth's climate. The theory of SR proposed for the first time that an appropriate additive random noise may enhance the probability of detection of a non-linear sub-threshold signal. Any system consisting of (a) non-linearity (through barrier or threshold), (b) a sub-threshold signal, and (c) additive noise with a proper variance is capable of exhibiting SR. There are many examples of SR in physical and biophysical systems such as, dithering system, Schmitt trigger, ring laser, Cray fish mechanoreceptor, cricket, human vision etc.

The idea of the association of noise with the nervous system is quite old. This led to the speculation of the positive role of noise in neural computation. It has been demonstrated in many experiments that the addition of external noise to a weak signal can enhance its detectability by the peripheral nervous system of crayfish [3], cricket [4] and also human [5-8] by the process of SR. In all these experiments the neural recordings were analysed, on the computer, for the presence of enhanced response through SR. All these were, therefore, indirect evidences of SR. It has also been demonstrated through psychophysical experiment [9] that human can make use of noise constructively for enhancing contrast sensitivity by the process of SR. It has been shown in this experiment that the brain can consistently and quantitatively interpret detail in a stationary image obscured with time varying noise and that both the noise intensity and its temporal characteristics strongly determine the perceived image quality.

It is well known that visual perception is a complex phenomenon involving higher level of cognition but it also includes lower level computation because for vision (visual computation) the very raw primal sketch is computed with the help of retina along with its associated circuitry. It is, therefore, expected that low-level computation (primal sketch) has a considerable role to play for observed enhanced visual perception by the process of SR.

We show in this work that a low-level visual computation performed by the center-surround receptive field along with the zero-crossing detection performed in the primary visual cortex together could be used to model the nature of contrast sensitivity enhancement, through SR, experimentally observed in human vision.

II. BACKGROUND

The human retinal network consists mainly of three layers of cells, a two-dimensional array of primary photoreceptors, a layer of bipolar cells and a layer of ganglion cells. Information from rods and cones are being sent to the bipolar cells, either directly or through the network of horizontal cells. The bipolar cells, in their turn, send information to ganglion cells, either directly or through the network of amacrine cells. Information from ganglion cells go to cortex through visual pathway. Investigations [10-11] revealed that the image is extracted in successive layers through a "centre-surround" effect. This 'antagonistic' center-surround effect is modeled by difference of Gaussian or DOG [12-13] for which the resultant looks like a Mexican hat in two-dimension.
 A DOG model in 2-D would be represented mathematically as:

---

[#] Corresponding author

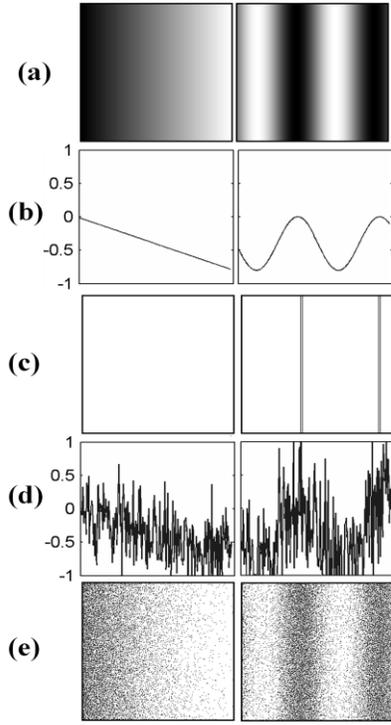

Fig. 1 Examples of zero-crossing maps of a ramp image and a sinusoidal grating (a) the original image ($I$), (b) profile of $I \otimes (-DOG - m\delta)$, (c) zero-crossing profile of the image in (b), (d) profile of $(I + noise) \otimes (-DOG - m\delta)$ and (e) zero-crossing map of the image in (d)

$$DOG(x,y) = A_1 \frac{1}{\sqrt{2\pi}\sigma_1} e^{\frac{-(x^2+y^2)}{2\sigma_1^2}} - A_2 \frac{1}{\sqrt{2\pi}\sigma_2} e^{\frac{-(x^2+y^2)}{2\sigma_2^2}} \qquad (1)$$

This model has been modified [14-15] to accommodate the concept of narrow channels [16] and the extended classical receptive field (ECRF)[17-20]. This is given by

$$-DOG(x,y) - m\delta(x,y), \qquad (2)$$

where $m$ is a constant factor given by $0 \leq m \leq 1$ and $\delta(x,y)$ is Dirac delta function in 2-D. There were claims of evidence in favour of zero crossing detector filters in the primary visual cortex [21-22]. This prompted us to use the above model for computing zero-crossing map of images for our investigation. The zero-crossing map, computed by (2), is capable of retaining shading information in the sense of stochastic halftone process as shown in Fig. 1(e). It may be noted that even though a gray level image after zero crossing would be converted into a binary picture, having only two gray values for all the points, namely, either 0 (totally black) or 1 (totally white), the shading information of the original image is retained in the zero-crossing map.

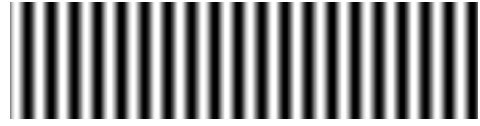

Fig. 2. Image of a typical sinusoidal grating, generated with $A\sin(2\pi f x) + 0.5$ using $f = 0.005$ cycles/pixels, used for the simulation study is shown.

### III. THE EXPERIMENT

To begin with we are investigating the usability of centre-surround model of retinal ganglion cells given by (2) for building a computational model capable of exhibiting SR. We would try to explain, with this model, some of the observations [9] related to the enhancement of contrast sensitivity with noise strength. Firstly we will show that the information content in the zero-crossing map of a sinusoidal grating of a given contrast, computed with (2), could be maximised in the presence of non-zero additive Gaussian noise. And secondly we will try to demonstrate, through simulation, the nature of the SR phenomenon experimentally observed in the contrast sensitivity of human vision [9].

The image for our investigation is a sinusoidal grating as shown in Fig. 2. The methodology of the simulation experiment is detailed below:

(a) We start with a synthetic image $I$ as shown in Fig. 2 generated by $A\sin(2\pi f x) + 0.5$ and digitized on a 0-1 gray scale. Here, the amplitude $A$, denotes the Michelson contrast of the picture, $f$ is the spacial frequency and x is the spatial coordinate along which the pattern is changed.

(b) A random number $n$ within 0-1, from a Gaussian distribution with zero mean and standard deviation $\sigma$ is added to original gray value $I$ in every pixel so that every pixel value becomes $I + n$. Thus the noise in each pixel is incoherent with that of all other pixels but the standard deviation remains the same for all.

(c) Next we choose a derivative filter function as in (2) with $f$ as the center frequency of the $DOG$, $A_1 = A_2 = 1$, $\sigma_2 = 2\sigma_1$, $m = 0.5$ and the special extent along x and y is taken as $10\sigma_2$ [23]. We compute the derivative of the image by $(I + n) \otimes (-DOG - m\delta)$ where $\otimes$ denotes convolution. The resulting derivative image becomes bipolar and the pixels will have negative as well as positive values.

(d) The zero-crossing map is then constructed from the resultant image in (c) by assigning a grayscale value 0 to each zero-crossing point and all other pixels in the image are assigned a value 1. This binary zero-crossing map resembles a stochastic halftone image where intensity variation of the original image maps to density variation of zero-crossing points. A typical example is shown on the right in Fig. 1(e).

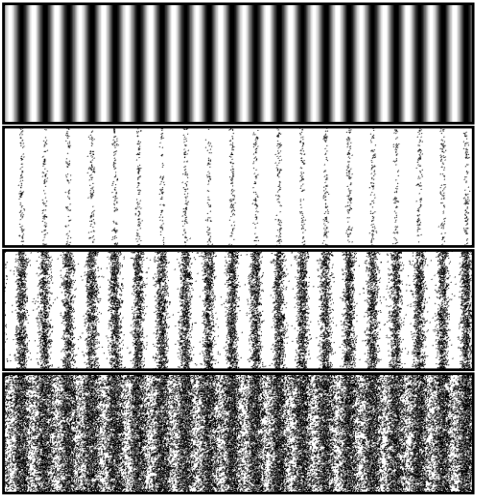

Fig. 3. Example of the effect of noise and contrast on the zero crossing is presented. The topmost one is the original sinusoidal grating and the lower ones are the zero crossing images for increasing contrast for a given noise.

(e) The steps (b)-(d) are repeated for various values of the contrast $A$ for a given noise strength $\sigma$. Typical example images for varying contrasts are shown in Fig. 3. The top most one is the original image and the rest of the images from top to bottom are zero-crossing images for increasing values of $A$. Even a visual inspection shows that the best reproduction of shading information is achieved (third picture from the top) with moderate a contrast value of $A$. This is a typical signature of SR.

(f) The next step is to have an estimate of the optimal contrast ($A_{opt}$) that will give best reproduction of the

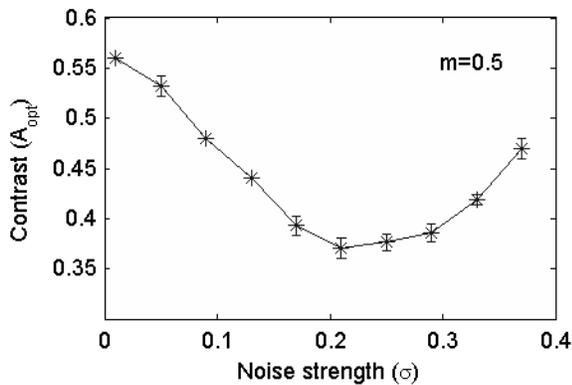

Fig. 4: Plot of optimum contrast with noise strength. The curve shows that the optimum contrast is minimum for an non-zero amount of noise.

shading information of the original image. Taking into account the observations made by [24] we study the zero-crossing image in the Fourier domain and look for the minimum contrast for which the second harmonic of the zero-crossing image just begins to appear. We designate this threshold contrast as $A_{opt}$. This is the optimal contrast for the noise strength $\sigma$ for which the second harmonic of the zero-crossing image just begins to appear. This will vary for different values of the noise strengths.

(g) We now repeat (a)-(f) for evaluating $A_{opt}$ for various values of the noise strengths. Archetypal behavior of $A_{opt}$ with $\sigma$ is plotted in Fig. 4. It is evident from the figure that threshold contrast $A_{opt}$ is the minimum for an optimal amount (non-zero) of noise strength and increases for all other noise strengths. Alternatively, contrast sensitivity ( $1/A_{opt}$ ) attains a maximum in the presence of an optimal amount of non-zero noise.

The above study shows that model represented by (2) along with the zero-crossing detection mechanism in visual cortex is capable of producing SR phenomenon in the presence of appropriate amount of noise when the input is a sinusoidal grating.

Our next agenda is to explore the applicability of this computational framework for explaining the phenomenon of noise enhanced contrast sensitivity in human vision observed in a psychophysical experiment [9]. For this psychophysical experiment Simonotto et al used the picture shown in Fig. 5(b) where the frequency varies along the spatial coordinate. Varying amount of noise was added to this picture of a given contrast, threshold filtered and presented to the subjects. The subjects were asked to identify the picture where they could no longer visually find a specified structure. From the feedback of the subjects the authors find the variation of their threshold contrast with noise strength which showed a typical signature of SR. This experiment can be thought of computationally similar to the simulation experiment presented above. The only major difference was that in the case of psychophysical experiment the noise contaminated images were threshold filtered before being presented to the subjects where as in the case of simulation no threshold were applied. We will, therefore, repeat the steps (a) – (g) with threshold filtration applied after step (b). We will also make one more change for the simulation. Taking into account the observation by the authors in [24] that contrast threshold of a grating is only determined by the fundamental Fourier component of its waveform, we propose to repeat the simulation with a sinusoidal grating whose frequency equals the fundamental component of the image in Fig. 5(b). The result of the simulation is presented in Fig. 6 for two values of the threshold $th$. The solid curve, taken from the threshold SR theory [9], is given by

$$A_{opt} = K\sigma \exp[th^2 / 2\sigma^2] \qquad (3)$$

and was fit to the simulated data (*) with $K$ as the only adjustable parameter. The quality of the fit to the simulated is very good.

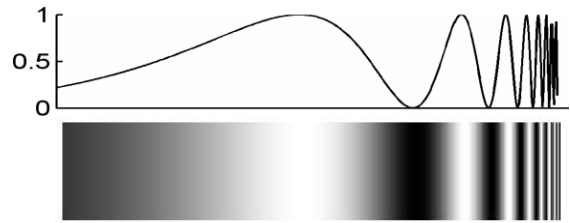

Fig. 5. A typical image used by Simonotto et al [9] for the psychophysical experiment is shown. These pictures were generated by $A\sin(1/x)+0.5$. (a) Plot of the function $A\sin(1/x)+0.5$ and (b) The image generated by the function.

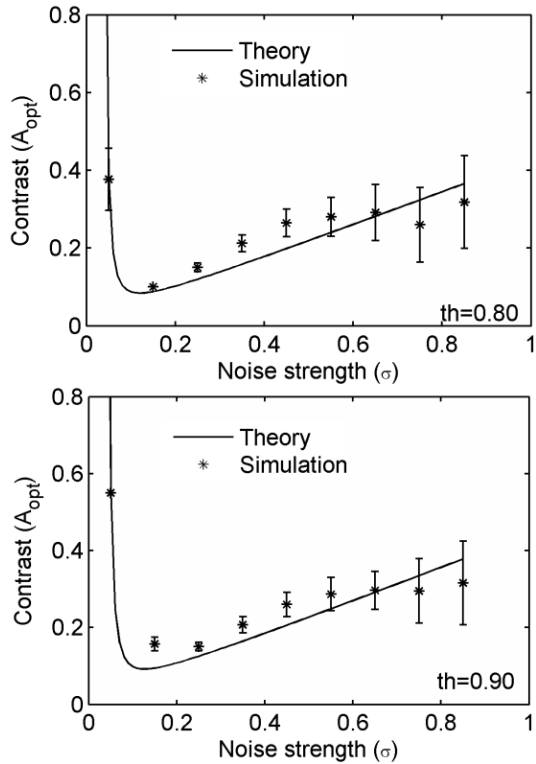

Fig. 6. Plots of the variation of threshold contrast $A_{opt}$ with noise strength for two different threshold values. The solid line is the fit of (3) with the simulated data.

## IV. RESULTS AND CONCLUSION

These experiments have demonstrated the utility of center surround model (2) for simulating some aspects of the visual system and its information processing in the presence of noise. The model could reproduce the nature of the enhancement of contrast sensitivity in the presence of optimal noise. The quality of the fit of the theoretical function (3) to the simulated data is surprisingly good. The repeatability and stability of the model suggests that it may become a useful tool for understanding how our visual system interprets fine detail within noise contaminated images. This can also be used to study and build artificial system for enhancing or repairing contrast sensitivity in human.

ACKNOWLEDGMENT

We are grateful to Subhajit Karmakar for stimulating discussions and important suggestions.